# Domain adaptation in small-scale and heterogeneous biological datasets

Seyedmehdi Orouji[1], Martin C. Liu[2,3], Tal Korem[3,4,5,*,†], Megan A. K. Peters[1,5,6,*,†]

**Author affiliations**
1 Department of Cognitive Sciences, University of California Irvine, Irvine, CA
2 Department of Biomedical Informatics, Columbia University Irving Medical Center, New York, NY
3 Program for Mathematical Genomics, Department of Systems Biology, Columbia University Irving Medical Center, New York, NY
4 Department of Obstetrics and Gynecology, Columbia University Irving Medical Center, New York, NY
5 CIFAR Azrieli Global Scholars Program, CIFAR, Toronto, Canada
6 CIFAR Fellow, Program in Brain, Mind, & Consciousness, CIFAR, Toronto, Canada

(* These authors contributed equally to this work.)
(† Corresponding authors. Email: tal.korem@columbia.edu, megan.peters@uci.edu)

# Abstract

Machine learning techniques are steadily becoming more important in modern biology, and are used to build predictive models, discover patterns, and investigate biological problems. However, models trained on one dataset are often not generalizable to other datasets from different cohorts or laboratories, due to differences in the statistical properties of these datasets. These could stem from technical differences, such as the measurement technique used, or from relevant biological differences between the populations studied. Domain adaptation, a type of transfer learning, can alleviate this problem by aligning the statistical distributions of features and samples among different datasets so that similar models can be applied across them. However, a majority of state-of-the-art domain adaptation methods are designed to work with large-scale data, mostly text and images, while biological datasets often suffer from small sample sizes, and possess complexities such as heterogeneity of the feature space. This Review aims to synthetically discuss domain adaptation methods in the context of small-scale and highly heterogeneous biological data. We describe the benefits and challenges of domain adaptation in biological research and critically discuss some of its objectives, strengths, and weaknesses through key representative methodologies. We argue for the incorporation of domain adaptation techniques to the computational biologist's toolkit, with further development of customized approaches.

# 1. Introduction

In the computational biological sciences, we are interested in learning informative "truths" about biological systems through machine learning or similar quantitative modeling techniques(*1*). Contrary to "purely statistical" correlations, we expect such "truths" to generalize beyond a specific dataset or population, indicating that they offer a grounded biological meaning. However, collecting (and sometimes labeling) biological datasets is difficult, expensive, and time-consuming, leading to many small but related datasets which are collected from different sources and under different environmental and experimental conditions (e.g. different labs, equipment, settings, humidity, etc). For example, in the widely used Autism Brain Imaging Dataset (ABIDE), functional magnetic resonance imaging (fMRI) data was collected at multiple sites, which hindered the ability to directly aggregate data(*2*). Beyond creating challenges in data curation and metadata standards(*3, 4*), this variability in the sources of small biological datasets creates different *domains* of data that have different statistical distributions.

While this variety is a strength that can facilitate discovery of generalizable truths, it also presents a significant challenge to computational biology: Applying knowledge gained from one dataset (a *source*) to another (a *target*) will fail if the two datasets possess highly divergent distributions – a phenomenon known as *domain shift* or *data bias(5, 6)*. In short, we cannot blindly apply a model (of any kind) trained on a source dataset collected under one set of conditions to new target data and expect it to perform effectively. In an age of open datasets and keen interest in adhering to FAIR principles (Findability, Accessibility, Interoperability, and Reuse of digital assets) to accelerate scientific discovery, it is increasingly urgent that we acknowledge the strengths and challenges of combining datasets.

To best extract generalizable insights while making use of *all* collected data from varying sources – especially in biological disciplines where data are expensive – and to apply these insights to newly collected data, we must discover how to best leverage the use of all existing and continuously growing small biological datasets(*7*). In the field of machine learning, *transfer learning* aims to use knowledge gained from learning a task on one dataset to performing a similar task on a different but related dataset, with the purpose of transferring knowledge across datasets(*8–12*). *Domain adaptation* (DA), a subfield of transfer learning, has been developed to address this issue of different statistical distributions by aligning the distributions of the source and target domains. Of note, while there are some similarities to "batch correction" often applied in high-throughput molecular measurements(*13, 14*), the objective is different: domain adaptation aims to learn generalizable models across domains, while batch correction is primarily aimed at removing technical variation. Importantly, DA is more than just "lining up the features" and training a model on both datasets; not only is this often impossible to do (especially if features are unlabeled), but statistical differences between the domains can often guarantee that such a brute force aggregation is doomed to failure. Instead, through DA, a model is forced to learn *domain invariant* features, i.e. features that are common across all domains, such that the learned model can be generalized and perform relatively well on a separate target domain. Another benefit of DA is that the integration of multiple datasets effectively increases the sample size, allowing for improved inference of statistical signals. This allows better use of available data and resources, reducing the need to collect and annotate expensive data(*15–17*).

However, using DA methods to extract informative and generalizable insights from different datasets is difficult in general, and is particularly difficult in computational biology. Compared to datasets typically used to train machine learning models(*18–21*), many "biological-scale" datasets are smaller in sample size, have many more features than samples, and have a complicated feature space (e.g. different numbers of features in each dataset, missing values, etc.). Therefore, while developing effective DA techniques that can work well with these small "biological scale" datasets to find general truths about biological systems is highly desirable, it presents a specific set of challenges to machine learning research.

In this Review, we aim to critically discuss the benefits and challenges of applying current DA methodologies and frameworks to such biological datasets. To this end, we use the token examples of functional magnetic resonance imaging (fMRI) and microbiome datasets, two seemingly different disciplines in biology, to show the common considerations critical to developing effective DA techniques in such data. Our goal is to lay out the key components that require consideration in selecting an effective DA technique, and highlight important areas of future methodological research in DA methods that can be maximally effective in biological datasets – especially as data sharing and metadata curation continues to mature.

## 2. Domain adaptation: a powerful tool for biological data

In the biological sciences – and especially as re-analysis and meta-analysis is facilitated through open data sharing – researchers often work with multiple distinct datasets collected through various procedures and techniques. These datasets may contain unique idiosyncrasies that are specific to a dataset, and which may or may not necessarily offer any biological insights (for example, different MRI participants or scanners(*2, 22–24*), or different patient populations for microbiome profiling(*25, 26*)). Additionally, each dataset alone may have high feature dimensionality despite small sample size, and thus may be overfit by modern, state of the art models(*27–30*) – making it challenging to learn robust models that will generalize. These factors make it particularly attractive to apply DA to aggregate such biological datasets, reducing overfitting and facilitating the discovery of "generalizable truths". Before assessing the challenges of doing so, we would like to briefly examine three specific benefits of DA for biological research.

### 2.1 Mitigating small sample size and large feature space

Ideally, a successful approach in computational biology is to fit a model with few free parameters across many samples. However, complex biological systems often need to be modeled with many free parameters, while training samples remain quite few. This degree of model complexity in the face of insufficient training exemplars can reduce generalizability and increase the risk of overfitting, where a machine learning model fits the training data all too well but fails to generalize to new, unseen data (e.g., cross-validation fails). To address this issue, domain adaptation (DA) can be used to integrate multiple individual datasets to increase the number of training samples available. This approach helps to achieve two essential goals: it provides access to a larger and more diverse set of training data, thereby reducing the potential negative impact of having a large number of parameters, and it also encourages models to be more properly regulated so they can better extract true signals rather than being overly sensitive to noise. Increasing generalizability in this way can also support other benefits, discussed next.

## 2.2 Transferring knowledge

Beyond simply increasing the number of training samples available, DA can also be used to transfer knowledge across different biological contexts (different cells, tissues, organisms, individuals, ecosystems, in-vitro, and in-vivo) – assuming that domains share some commonalities in between features and task or goals. This could help scientists and physicians to transfer knowledge from some existing rich datasets to a different (but related) dataset that is smaller in size. For instance, in many situations there exists a large amount of labeled data from adults' MRI scans but much less data for infants; therefore, DA might be especially helpful to transfer insights gained from adults to newborns(*31*). DA could also help transfer drug response insights gained from richly annotated pre-clinical cell lines to more poorly annotated human settings(*32*), or to use DNA methylation data from multiple distinct tissues to predict donors' age(*33*). In general, it is highly desirable to transfer knowledge gained from existing labeled datasets to other different, but related, datasets that are sparse in terms of sample size or annotation. It is easy to envision the benefit of applying models trained on publicly-available data to a locally-collected, small dataset – a process made potentially much more powerful through DA.

## 2.3 Discovering generalizable patterns

As introduced above, DA can also help drive at our primary scientific goal: to reveal true, meaningful, and generalizable biological insights rather than associations that are merely due to artifacts, confounds, idiosyncrasies unique to one dataset, or meaningful biological differences between domains which are separable from a particular question at hand. This is crucial since biological datasets are often composed of many different small cohorts collected from different laboratories and under different environmental and experimental conditions(*20, 29*). For example, many fMRI(*34–36*) datasets are small, consisting of 30 human subjects or fewer per scanning site, but different hardware components or settings across MRI machines may result in data with different statistical distributions – e.g., different noise characteristics, signal magnitudes, correlations between features, or stationarity of these components across time for each scan site. In the microbiome field, the vaginal microbiome has been studied in over a dozen cohorts in the context of preterm birth(*37, 38*), and the gut microbiome has been similarly studied in the context of colorectal cancer(*39, 40*), yet variability in microbiome profiling across laboratories has been repeatedly noted(*25*).

As these smaller, individual datasets are increasingly shared and curated into large databases, challenges of discovering domain-invariant patterns while using *as much data as possible* become immediately apparent. Because of the idiosyncratic nature of each individual dataset, machine learning models can learn non-relevant information in one single 'training' dataset that can lead to incorrect *general* conclusions about biological processes. For example, even sophisticated computer vision models can discover 'shortcuts' when detecting COVID-19 from chest radiographs: Instead of detecting clinically relevant factors, they rely on confounding factors such as laterality markers or patient positioning. This not only hinders the ability of the models to generalize to new data (i.e., when tested on a new patient from a new hospital)(*41*) but also might lead to misinterpretation of results within a single dataset. This issue is related also to *batch effects(42)* – essentially, the effect of non-biological artifacts that changes the distribution of the data for an experimental subset of a particular experiment. When experimental batches (e.g., plates for DNA extraction, or days for MRI appointments) are also associated with the outcome of interest, it may even lead to incorrect conclusions (batch *confounding*). DA can be used to correct for these domain-specific idiosyncrasies when combining batches or cohorts, facilitating discovery of domain-invariant signals which may be more meaningful biologically.

# 3. Challenges of domain adaptation in bio-scale data and a path forward

Despite the clear utility of DA in biological data, its successful application to small datasets with complex features comes with significant challenges – many of which stem from the very reasons we would want to use it in the first place. In service of laying out a path forward to effective deployment of DA methodologies in biological scale datasets across multiple fields, we next explore in more detail why existing DA techniques may not be able to perform effectively on biological datasets. The purpose of this discussion is to help researchers learn to evaluate DA approaches for appropriateness in their own research, as well as to highlight deficiencies in current DA applications to biological questions which may be alleviated through improved collaboration between DA researchers and computational biologists.

## 3.1. Number of samples and features

Most DA methods have been designed in the fields of computer vision, text mining, or language processing(*43–46*) with reference to – and evaluation on – large-scale text and image data, where there can be tens of thousands (or even millions) of samples available for training (e.g. MNIST, CIFAR10; refs.(*47–49*)). In contrast, the number of samples in biological datasets is often small, but they simultaneously have many features, a problem known as curse of dimensionality(*50*). For instance, in a typical fMRI or microbiome dataset we might only have a few dozens to hundreds of samples while the number of features could exceed thousands(*26, 51, 52*). As introduced above, this imbalance between the number of samples and features can potentially lead to overfitting problems(*27, 28*), which in turn hinders the effectiveness of DA techniques on biological datasets(*50*). There do exist several datasets typically used to benchmark DA approaches that may be somewhat closer in size to biological-scale data, including Office31 (ref. (*53*)), which contains image data of objects collected from 3 source domains with different resolutions, for a total of 4,110 images from 31 object categories (132 images per category). However, while one might hope that DA methods that have shown success on Office31(*54–56*) could be useful for biological data with similar sample size per category, it must be acknowledged that many biological datasets have significantly different properties than imaging data(*57–60*), and are even smaller, with only several hundred training samples in total. There is a need for DA algorithm development to specifically target success in the face of fewer training samples.

## 3.2 Differences in feature complexity

However, simply checking that DA approaches can perform adequately on small datasets is unfortunately unlikely to be enough. Another barrier to applying DA approaches to biological data is that features in biological domains are inherently much more complex than those in image data. For example, in many machine learning datasets such as MNIST or Office-31, image data are essentially pixel luminance values in the RGB and alpha channels that can be relatively simple to aggregate with other source data, for example by resizing the image(*5, 61–64*). In the case of biological datasets however, the inherent complexity of features can significantly hinder our ability to aggregate different sources of data. For example, biological datasets often contain missing values(*65–68*), or have different numbers of features with unknown mapping orders between domains(*24*) (i.e., which features in a source are "the same" as which features in a target domain). They can also exhibit non-linear relationships or interactions between features(*29, 68–70*), and unique data preprocessing requirements for each source can substantially increase the complexity of developing DA techniques for biological datasets. In other words, in addition to feature-to-sample ratio and number of categories, we need to take into account the complexity and heterogeneity of biological domains before using DA techniques on biological datasets. This increased complexity stems from several sources which we next discuss in more detail.

### 3.2.1 Missing values

Biological samples often contain many missing feature values. For example, microbiome data typically only consists of a few taxa that are shared by the majority of samples, and even less so across cohorts. Many taxa are rare and are only detected in very low abundances, a phenomenon known as zero inflation in statistics(*71*). In human neuroimaging, PET or MRI scans combined with patients' genetic information can help with early diagnosis of Alzheimer's disease. However, the very common problem of missing values (i.e. not every subject has completed multi-modality data) can impede the ability of these multimodal models to make reliable predictions(*72–74*). Missing data is less problematic in many traditional datasets used to train DA approaches, meaning that these approaches may not deal with missing data well; to be successful with biological data, DA algorithms need to adequately handle small data and missing values.

### 3.2.2 Heterogeneity of features

Biological domains also often possess different numbers of features, and the features also often do not lie in the same rank order across domains. For example, fMRI data from a given brain region will have different numbers of voxels from one human subject to the next, and the information represented, for example, in voxel 1 in person A is unlikely to functionally align with the information encoded by voxel 1 in person B. While functional alignment approaches have been developed(*24*, *75*), they do not explicitly perform DA operations. In microbiome research, it can be unclear whether a particular taxa is the same across datasets, especially because sometimes the measurement techniques differ (e.g., taxa are characterized using different regions of a marker gene, such that the same taxa might be represented by different features in different datasets). These examples are in stark contrast to most image-based DA approaches, which can exploit physical proximity of features (pixels) through spatial convolution or learn feature importance maps based on spatial features alone (e.g., the center of an image may often be more informative than the edges).

Additionally, domains may have some overlapping features but also some non-shared (distinct) features – i.e., those that are specific to one domain but not the other(*76*). Current DA techniques may not be very effective on such datasets since domains may lack supplementary information such as labels(*77*) or information about matching features or samples between datasets(*11*). This limitation could force researchers to remove domain-specific features and hence lose the capacity of DA models to benefit from these unique features in the learning process. Ideally, DA for biology could benefit from a specific focus on both feature alignment (ideally unlabeled) and principled ways to deal with shared versus non-shared features.

### 3.2.3 Distribution of feature importance

In biological datasets, feature importance distributions can be more highly skewed than in many standard benchmarks used to test DA approaches. That is, in biology, a *few* features can be *very important* for the ultimate performance of a model; in contrast, in typical benchmark datasets, many features can have similar importances(*57–60*). This difference in skewness of feature importance distributions can lead to extreme challenges for many DA approaches, such that DA models which succeed even on small 'typical' benchmark datasets may fail in biological applications.

## 3.3 Contributions of data collection and preprocessing procedures

Biological datasets often require extensive preprocessing after the data collection stage which can be inconsistent across datasets or laboratories (DADA2 or deblur for 16S rRNA amplicon data(*78*, *79*), fMRIPrep(*80*) versus AFNI(*81*, *82*), or FSL(*83–85*) for fMRI images(*86*)). As a result, machine learning methods used in biology typically are limited to being highly context- and preprocessing-specific, requiring careful design and tailoring to test the desired hypothesis appropriately(*87*). This often occurs despite targeted efforts in bridging this gap by the means of setting up standards in generating and preprocessing the data(*88*), since some lab- and individual-specific idiosyncrasies are wholly unavoidable. For example, in fMRI data correction for subject's head movement, using different scanning sequences or scanners can introduce data shifts that makes applying DA techniques even more difficult(*2*, *89–93*). Such preprocessing idiosyncracies can exacerbate or interact with other batch effects, including introducing or altering interdependencies among features(*29*).

## 3.4 Interpretability of features and feature spaces

Interpretability is an important aspect of biological research, in contrast to at least some other ML applications. However, alignment steps in DA – which often require finding a latent representation of data by projecting the domains into a shared feature space(*94–96*) – are frequently carried out by machine learning and deep learning methods. This means that DA in biological data inherits the same problem that plagues machine learning more broadly: failures in interpretability due to the black-box nature of these machine learning and deep learning methods. In fact, the shared feature space is particularly challenging to interpret(*97*) because the shared feature space is defined as a latent space that bridges two or more domains, rather than the latent space defined by one domain alone. Therefore, DA research can and should aim particularly at understanding how input features are related to the common feature space when utilizing these methods(*98*, *99*).

## 3.5 Theoretical limitations of domain adaptation

It is also important that we discuss a critical theoretical limitation of DA, especially as it might impact biological data. The primary driver of DA's potential success is the *adaptability* between the source and target domains(*100*, *101*) – essentially, the theoretically maximal ability of an ideal model to jointly model them(*102*, *103*). Failure of adaptability is thus a potentially fatal concern. While considering additional source domains provides the benefits of a larger and more diverse sample set (or additional labels), these domains might have inherently different distributions of features or different joint distribution with the labels, which would make a model considering them less accurate. Thus, applying DA might ultimately bring more cost than benefit(*101*): in the worst-case scenario, *negative transfer* (i.e., applying knowledge from a source domain negatively affects the performance of the model in a target domain) can happen(*100*, *104*). Crucially, the potential for negative transfer can be further amplified when working with biological data, due to its already-heterogeneous nature and the smaller sample size of each dataset. Therefore, it is imperative that the adaptability of the particular biological datasets in question be explicitly quantified or estimated before applying DA methods. Unfortunately, while there exist a few methods to quantify adaptability between domains(*100*, *102*), analysis in the context of different biological sub-fields is exceedingly rare. The development of adaptability analysis methods thus may be a fruitful and critical area of future research into DA application to biological datasets.

## 4. Considerations for domain adaptation

Despite the challenges noted above, even in their current state, DA approaches can still provide benefit in biological data at this critical expansion of data sharing and open science practices in biology. But there are a great many methods to choose from. How should a scientist select the best DA approaches for their own datasets or scientific questions? In this section, we outline specific considerations for biologists in selecting and applying DA approaches in their own research.

We begin this section by presenting a formal definition of *domain* and *domain adaptation*. We then present a taxonomy which can be useful in gaining a better understanding of what to search for in the literature. In this Review we focus on the primary subcategory of DA which addresses *data bias* or *covariate shift*; this DA subcategory tries to align shifts in the feature spaces between domains (or the change in the marginal distribution of data samples across domains). Other specialized subcategories of domain shift include *label shift(105)*, which indicates that different domains contain different number of labels for each class, and *concept shift(106)*, in which the data distribution remains the same but the conditional distribution changes (i.e. $P_s(y|X) \neq P_t(y|X)$). Interested readers should refer to these surveys(*107*, *108*) for a comprehensive overview of the different types of shifts in the DA field.

## 4.1. What is a domain?

A domain can be defined as $D = \{\chi, P(X)\}$, where $\chi$ is a feature space, $X = \{x_1, x_2, ..., x_n\}$ is an instance set with $x_i$ denoting a given feature, $n$ denotes the number of features or dimensions in the data (e.g., in fMRI data voxel activities or taxa in microbiome data), and $P(X)$ denotes the marginal probability distribution of all samples in that dataset. This formal definition is typically used in discussions of DA across a wide variety of disciplines(*109*, *110*).

## 4.2 The terminology of domain adaptation

For a specific domain, we define the task (e.g., predicting what image a subject is looking at from neuroimaging data, or predicting a disease state from microbiome composition) as $T = \{y, f(\cdot)\}$, where $y$ denotes the labels to be predicted and $f(\cdot)$ denotes a decision function (i.e., the posterior probability distribution of $P(y|X)$ of the joint distribution $P(X, y)$) that needs to be learned in order to map input features to the corresponding labels.

Given these definitions, domain adaptation is faced with the following problem, in which distributions or relative alignment of features across domains are different but the task remains approximately the same. Thus, a DA problem with covariate shift can be formally defined as follows:

$$P(X_{s_1}) \neq P(X_{s_2}) \neq ... \neq P(X_{s_k}) \neq P(X_t),$$
$$T_{s_1} \approx T_{s_2} \approx ... \approx T_{s_k} \approx T_t$$

where $s$ denotes the source domain, $t$ denotes the target domain, $k$ is the number of source domains, $P(X)$ is the marginal distribution of a specific instance set in a given domain, and $T$ is the task performed in each domain. Here, the goal of DA is to improve the performance of target decision function $f(\cdot)_t$ in target domain $D_t$ by leveraging the information from source domain $D_s$ and decision function $f(\cdot)_s$ (which is learned on the source domain after the source and target domains are aligned). In other words, DA intends to adapt the model(s) trained from a source (or sources) to a different, but related, target dataset. It does this by aligning the distributions of features and samples belonging to different domains so that the models emphasize learning *domain invariant* features that are not dependent on a specific dataset (**Figure 1**).

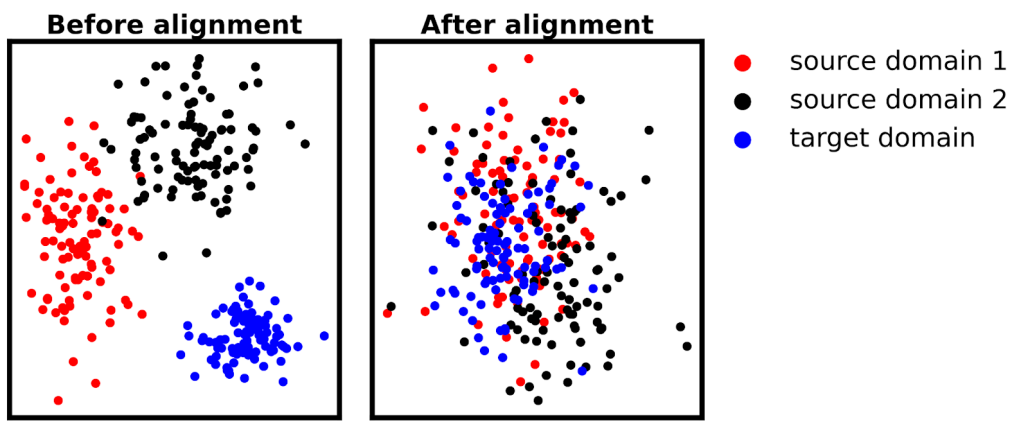

**Figure 1.** A cartoon representation of source and target domains before and after alignment. In this cartoon, features vary in their values along two dimensions, and each domain's features take on a different mean and covariance. Unless the domains are aligned, these differences could both obscure other meaningful variation in the data that is shared across domains, and prevent models trained on one domain from generalizing to another.

## 4.3 A taxonomy of domain adaptation

Generally, when undertaking a DA analysis, we should consider three main factors:
1. The data used to train a model may be collected from **multiple sources** or just from a **single source**.
2. Depending on the availability of labels in the target domain, we might choose **supervised**, **semi-supervised**, or **unsupervised** models.
3. The feature spaces in the source(s) and target domains can be **homogenous**, meaning that they have the same dimensionality and "meaning", e.g., feat4–0w5ure A in source 1 represents the same "type" of information as feature A in source 2; or **heterogenous**, meaning that the feature spaces may differ in terms of dimensionality and/or meaning.

In the following, we discuss these three factors in more detail. **Table 1** also shows a summary of these categories accompanied by mathematical annotations.

### 4.3.1 Single- vs. Multi-source

In selecting a DA method, one question you will want to ask is how many domains are present. As mentioned above, DA techniques can be divided into two categories of "single-source" and "multi-source"(*111*). In single-source DA, the source domain is usually labeled, while the target domain belongs to another domain that possesses a different distribution(*12, 96*). Single-source DA is simpler than multi-source DA since there are only two distributions of data – source and target. Therefore, single-source DA is a good technique when there is enough data available in both the source and target domains to effectively train a model that can perform well on the target domain(*112, 113*)(*114, 115*).

However, in modern real-world data sharing initiatives, most biological data come from many sources(*116*, *117*), and using this data to its full extent can facilitate novel insights. Therefore it is advantageous to find models that leverage all available sources. This problem can be addressed through multi-source DA, which aims to combine multiple sources of labeled data to make predictions about a similar task on a target dataset(*111*, *116*, *118*, *119*). A naive way to solve this problem is to combine multiple sources into one big source domain and then approach the problem as a single-source DA(*111*, *120*). However, these methods can show very limited improvement in performance – and sometimes even worse performance – in comparison to using only one source(*121*), specifically stemming from challenges of aligning the sources to begin with. Another way to tackle this problem could be to train a model on each source independently, apply each trained model to the target domain, and then vote for the 'correct' label in the target domain based on the prediction across sources(*122*). One could also attempt to first discover domain-invariant features among all source and target domains(*123*), or use a two-stage alignment technique that first tries to find domain-invariant feature spaces for each source-target pairing and then align model outputs across these spaces(*121*). In all cases, though, Multi-source DA is significantly more challenging than single-source DA – a problem made worse by the particular characteristics of biological data, as discussed above.

### 4.3.2 Supervised vs. semi-supervised vs. unsupervised

It is also important to assess what kinds of labels are available for your data, across all the domains you need to align; this will dictate whether you should select a *supervised*, *semi-supervised*, or *unsupervised* DA method. These labels have been applied in varying ways(*12*, *111*, *124–126*). Here we have chosen a categorization based strictly on the usage of target labels: in unsupervised DA, no label is available in the target domain(*55*, *96*, *127*, *128*); in semi-supervised DA(*129–131*), some labels are available to use; and in supervised DA, labels in the target domain are available for most samples(*107*). Although the majority of DA techniques in existing literature focus on unsupervised DA (since it is often utilized for the purpose of annotating unlabeled data in the target domain), in the case of biological data, any of the supervised, semi-supervised, or unsupervised scenarios is possible. This is because the primary goal of domain adaptation in biological settings is to uncover insights about biological systems that generalize across domains. Thus, even when labeled data are available in the target domain, one can still benefit from utilizing DA techniques on different datasets to find generalizable patterns across domains.

### 4.3.3 Homogeneous vs. heterogeneous

Finally, it is important to understand how the features are related across your different domains. DA can be divided into two categories based on the relationships between these features: homogeneous or heterogeneous(*107*, *109*, *111*). In homogeneous DA, the source and target domains have the same feature space, $\chi_s = \chi_t$, but the data distributions of instances of these feature spaces are different, $P(X_s) \neq P(X_t)$. That is, feature 1 in domain 1 represents the same "meaning" as feature 1 in domain 2 – for example, they both represent a specific voxel at a specific coordinate in the brain, or represent the same microbe (Note: $\chi_s = \chi_t$ means that the feature space in both domains is homogenous, but if $X_s = X_t$ then it means that $X_s$ and $X_t$ are identical datasets such that there is no difference between the source and target datasets at all). In heterogeneous DA, conversely, the feature space is related but different between the domains. Many DA techniques that have been developed so far tend to focus on homogeneous DA(*95, 132–140*). For instance, the source data could be the fMRI data obtained from a subject with one scanner and the target domain is the fMRI data obtained from the same subject with the same protocol but a different scanner. Alternatively, different domains could contain gut metagenomic sequencing data from different studies aligned against the same reference database. Addressing the domain shift in a homogeneous DA problem is relatively simpler since it is possible to simply perform the feature alignment directly on the original instances of the domains without the need to project them into a common feature space.

Unfortunately, however, most biological datasets are heterogeneous in nature(*29, 68*) since these data are collected in different laboratories, under different environmental and experimental conditions, and sometimes even for answering different but related questions. In other words, neither the feature spaces nor the marginal distributions are the same (i.e. $\chi_s \neq \chi_t$, $P(X_s) \neq P(X_t)$). As a result, biological datasets very often have different feature dimensionalities, and sometimes these features even have different labels or come from different modalities of data collection (e.g., fMRI versus another neuroimaging modality like electroencephalography). For instance, the fMRI data from the brains of two individuals have different numbers of voxels (features) which also are not meaningfully aligned across individuals regarding their functional properties (e.g., voxel 1 in person A is unlikely to encode the same information as voxel 1 in person B) – even when the scanner, protocol and performed task are exactly the same.

| | **Categories \| Definitions** | **Domains,** $D = \{\chi, P(X)\}$ **& Tasks,** $T = \{Y, f(\cdot)\}$ | **Verbal description** |
|---|---|---|---|
| **Traditional ML vs transfer learning** | **Traditional ML** | $D_s = D_t \ \& \ T_s = T_t$ | When the source (i.e. training set) and target (i.e. test set) have the same distribution and the task is exactly the same. |
| | **Transfer Learning (TL)** | $D_s \neq D_t$ or $T_s \neq T_t$ or both | When the source and target domains have different distributions or the performed task on source and target are different, or both. |
| **Single- vs multi-source DA** | **Single-source DA** | $P(X_s) \neq P(X_t)$ & $T_s \approx T_t$ | When there is only one source domain and the marginal distribution of the feature space between source and target domain is different. The task in the target domain is similar to that in the source domain. |
| | **Multi-source DA** | $P(X_{s_1}) \neq P(X_{s_2}) \neq ... \neq P(X_{s_k}) \neq P(X_t)$, & $T_{s_1} \approx T_{s_2} \approx ... \approx T_{s_k} \approx T_t$ | When there are multiple sources available which can have different distributions, and when these distributions differ from that of the target domain. The task is similar across all domains. |
| **Supervised, semi-, or unsupervised** | **Supervised** | $P(X_s) \neq P(X_t)$, *with all target labels* | When source and target domains are both labeled. |
| | **Semi-supervised** | $P(X_s) \neq P(X_t)$, *with some target labels* | When source is labeled but target is partially labeled. |
| | **Unsupervised** | $P(X_s) \neq P(X_t)$, *with no target labels* | When source is labeled but target is not labeled. |
| **Homogeneous vs heterogeneous** | **Homogeneous DA** | $P(X_s) \neq P(X_t)$ & $\chi_s = \chi_t$ & $T_s \approx T_t$ | When the feature spaces have the same dimensionality and same meaning. |
| | **Heterogeneous DA** | $P(X_s) \neq P(X_t)$ & $\chi_s \neq \chi_t$ & $T_s \approx T_t$ | When the feature spaces have different dimensionality or different meanings. |

**Table 1. Difference among traditional machine learning, transfer learning, and various kinds of domain adaptation.** ML, machine learning; DA, domain adaptation. χ represents

feature space, and $P(X)$ is the marginal distribution of instance set $X$, $T$ denotes the performed task, and $f(\cdot)$ is the decision function to map each sample to the corresponding label. $s$ denotes the source domain, $t$ denotes the target domain, and $k$ is the number of source domains.

## 4.4 Case studies and practical examples

Given the nature of most biological datasets, which often contain limited samples and originate from many different sources, the most common DA setting in this field is multi-source heterogenous DA settings. For instance, aggregating fMRI data from multiple subjects or even multiple sites(*34–36*) can be considered a multi-source heterogeneous domain adaptation. It is multi-source because the data is coming from multiple subjects or multiple sites with different MRI scanners, and it is heterogeneous because the number of voxels (i.e. features) from each subject and the information they represent is different. (Note: number of voxels can be equated through spatial normalization to a standardized template, but this does not address that each voxel will still represent different information across individuals.) In the microbiome field, integration of data from multiple microbiome datasets in order to predict a phenotype on a held-out study(*37, 38, 141*) is once again multi-source and heterogeneous, as data are often amplicons of different regions of the 16S rRNA gene. To illustrate the utility of existing DA approaches and explore their categorization with the taxonomy discussed above, here we select several methods to discuss in slightly more detail.

One DA method, the PRECISE method(*32*), has been used to predict patients' drug response based on available pre-clinical datasets such as cell lines, and patient driven xenografts (PDXs). To achieve this, the authors first extracted factors from cell lines, PDXs and human tumors using principal component analysis (PCA). Then they aligned these subspaces from human tumor data with pre-clinical data using geometric transformations, and extracted common features associated with biological processes followed by training a regression model using consensus genes and validated with known biomarker-drug associations to accurately predict drug response in patients. In this study, DA was homogenous, as the features (genes) in the source and target domains were the same; multi-source, as various source domains were used (i.e. cell lines); and supervised, as the labels of all samples were used.

Another method, Adversarial Inductive Transfer Learning (AITL)(*142*), similarly aims to utilize largely available source domains such as cell lines and clinical trials to predict drug responses on small and hard-to-obtain gene expression data from patients. To this end, researchers first used a feature extractor network to map the source and target into a common feature space. This mapping aimed to alleviate the domain shift by using a global discriminator to learn domain-invariant features. Then, these domain-invariant features were used to build a regression model for the source task (i.e. predicting IC50) and a classification network to make predictions on the target task (i.e. predicting whether there is reduction in the size of the tumor). This study aimed to address both prior and covariate shifts in the source and target domains. The data used in this study came from multiple heterogeneous sources including thousands of cell lines from different cancer types. Finally, the target samples were labeled. This study can thus be characterized as a multi-source and supervised heterogeneous (i.e. drug response is categorized differently between preclinical and clinical settings) DA scenario.

Other methods such as WENDA(*33*) (Weighted Elastic Net for unsupervised Domain Adaptation) aim to predict a human's age using DNA methylation data, which are known to be different across different tissues. WENDA aims to use the available DNA methylation data from some tissues (source domains) to predict the age of the human subject using DNA methylation from a different tissue (target domain) by giving more importance to features that are more robust and behave in a similar fashion across source and target domains. In this study, data from 19 different tissues with chronological age ranging from 0 to 103 years old were used as the source domain. The target domain came from 13 different tissues, with chronological age ranging from 0 to 70 years old. In the application of WENDA, the source domain remained unchanged, while each tissue type was viewed as a distinct target domain. This thus represents a multi-source, unsupervised, homogenous DA scenario.

In another study, Li and colleagues(*2*) propose a multi-source domain adaptation approach by using resting-state fMRI "Autism Brain Imaging Data Exchange" (ABIDE) datasets(*143*) from multiple academic sites (UMI, NYU, USM, UCLA). Their goal was to improve the classification accuracy of autism diagnosis by detecting biomarkers. In this study, the feature space, denoted as $\chi$, was extracted features from fMRI sites such that $\chi_i = \chi_j$, with *i* and *j* representing different institutions (the data can be spatially normalized across participants by warping to MNI space). From this perspective, this problem is a homogeneous domain adaptation scenario. Subsequently, the authors utilized a Mixture of Experts (MoE)(*144, 145*), combining multiple neural networks – each of which is specialized in solving a specific task – in order to improve the overall performance of the model, and adversarial domain alignment methods to minimize the discrepancies between the domains, and successfully demonstrated the advantage of using federated domain adaptation techniques in using multi-site fMRI dataset to classify autism. Additionally, they were able to reveal possible biomarkers in the brain for autism classification. Therefore, in this framing this can be considered as a multi-source and supervised homogeneous DA problem.

Finally, Gao and colleagues proposed the deep cross-subject adaptation decoding (DCAD)(*146*) method: a single source, unsupervised, heterogeneous domain adaptation technique. DCAD uses a 3D feature extraction framework using 3D convolution and pooling operations based on volume fMRI data to learn common spatiotemporal patterns within a source domain to generate labels(*146*). Subsequently, an unsupervised domain adaptation method minimizes the discrepancy between source and target distributions. This process considers different subjects as different sources and aids in the precise decoding of cognitive states (in working memory tasks) across subjects. To validate the approach, they applied task-fMRI (tfMRI) data from the HCP(*147*) dataset. The experimental outcomes revealed exceptional decoding performance, achieving state-of-the-art accuracy rates of 81.9% and 84.9% under two conditions (4 brain states and 9 brain states, respectively) during working memory tasks. Additionally, this study demonstrated that unsupervised domain adaptation effectively mitigates data distribution shifts, offering an excellent solution to enhance cross-subject decoding performance without relying on annotations.

## 5. Future directions

### 5.1 What is missing from DA approaches in biological applications?

Despite these exciting successes, continued development of DA approaches tailored to the challenges of biological data is critically needed. This is especially important in light of the increasing availability of curated open datasets, complemented by increasing standardization of metadata standards(*3*, *4*). We thus hope the machine learning community will continue to develop techniques that can address relevant limitations of biological datasets, including:

1. Models must be able to capture the non-linear and complex patterns in biological systems, ideally with minimal or no assumptions. Therefore, many linear-based domain adaptation techniques (usually focused on some sort of transformation from source to target domain) might not be adequate.
2. Ideally we want to utilize domain adaptation to discover the underlying mechanisms of biological phenomena, rather than simply aggregating data for automatic annotation. Unfortunately, many existing techniques are primarily developed for addressing automatic annotation of unlabeled data. Therefore, to fully unleash the power of DA in biological systems, we must focus on methods that seek to discover domain-invariant features that are common across datasets. This usually happens by mapping all domains into a common feature space.
3. This domain-invariant mapping should be done using methods that work with limited data in individual cohorts. Although deep learning models are great tools to uncover highly nonlinear and complex relations in data with no specific assumptions, they often require many samples. Recently, simpler neural network architectures such as TRACE(*148*) and Fader networks(*90*) have shown promise with small fMRI datasets. However, many of the powerful neural network architectures such as GANs might not be suitable for biological datasets as they usually require vast amounts of data(*149*, *150*).
4. Methods should be developed to address domains' *adaptability* with specific focus on biological datasets. As mentioned earlier, methods do exist to quantify adaptability between domains(*100*, *102*), but limited attention has been paid to how such methods may fare in biological contexts.

In sum, it is incumbent upon us in the biological disciplines to challenge machine learning research to design more flexible and broadly applicable DA methods that can perform under the constraints of real-world biological datasets. An important step towards this goal will be to test and evaluate existing approaches on our own data, and on data available through increasingly comprehensive and consistently annotated shared data repositories, to comprehensively explore and categorize their current shortcomings. Thus, we hope that, with the help of the topics discussed in this Review, researchers in biological disciplines will feel empowered to try out existing DA approaches and to help catalog their successes and shortcomings.

If you would like to use DA techniques to augment your own data processing pipeline, we urge you to begin by gaining a comprehensive perspective on your data using the definitions and taxonomy described above. For example, How many sources do you have available? What is the sample size in each source? Do these sources contain equal amounts of features? If not, what are the nature of features in each source? Are these features in each source known and have a label? What task are you trying to achieve? Depending on the answers, you can choose the appropriate DA approaches, and set about examining their successes or failures. We hope that the tools and information provided in this Review will encourage you to do so, and to report your findings so that iterative improvements in DA approaches can be made to best serve our fields.

## 5.2 Promises for the future

In this piece we have focused on human neuroimaging (specifically fMRI) and microbiome sciences as token examples to speculate the potential promises of DA in computational biology as a whole. We hope that these selected case studies have helped to show off the potential of DA in numerous and varied biological disciplines, from electrophysiology, multi-omics, DNA sequencing, and scRNA sequencing to and protein localization – all of which face similar challenges in data collection and labeling to the case study fields discussed here. Differences in equipment, experimental setup, or even individuals can lead to a shift in the distribution of data, even when the task is identical. In all cases, however, our goal as researchers and clinicians is to go beyond domain-specific or dataset-specific models in order to discover domain-general and informative "truths" about biological systems.

Thus, DA could be extremely useful to aggregate diverse biological datasets available across the Open Science Framework, OpenNeuro, Neurosynth, Dryad, CEDAR, and more in search of meaningful and even clinically relevant outcomes(*151–154*). But much work is needed to address the existing challenges. It is the intention of this paper to help and facilitate these processes by bringing more awareness of DA, and the need to develop new techniques that are compatible with the limitations of biological datasets in order to make it accessible to biologists. If we are successful in identifying the challenges of performing DA on biological data, we are optimistic that DA and transfer learning methodologies can greatly benefit biologists.

## Author contributions

**Seyedmehdi Orouji:** Methodology; Investigation; Data Curation; Writing - Original Draft; Writing - Review & Editing. **Martin C. Liu:** Methodology; Investigation; Data Curation; Writing - Original Draft; Writing - Review & Editing. **Tal Korem:** Conceptualization; Methodology; Investigation; Resources; Data Curation; Writing - Original Draft; Writing - Review & Editing; Supervision; Project administration; Funding acquisition. **Megan A. K. Peters:** Conceptualization; Methodology; Investigation; Resources; Data Curation; Writing - Original Draft; Writing - Review & Editing; Supervision; Project administration; Funding acquisition.

## Acknowledgements

We thank Harry H. Lee for initial discussions on this topic. This work was supported by two Canadian Institute for Advanced Research Azrieli Global Scholars Fellowships (in Brain, Mind, & Consciousness [to MAKP] and in Humans and the Microbiome [to TK]), and a Canadian Institute for Advanced Research Catalyst Grant (to MAKP and TK). The funders had no involvement in the design or content of this work.

## Competing interests statement

The authors declare no competing interests.

# Domain adaptation in small-scale and heterogeneous biological datasets - Supplementary Information

Seyedmehdi Orouji[1], Martin C. Liu[2,3], Tal Korem[3,4,5,*,†], Megan A. K. Peters[1,5,6,*,†]

Here we provide more detail on specific domain adaptation methods than offered in the main text. The purpose of this section is to provide a greater degree of detail for the interested reader, as well as to suggest further reading through the works cited.

## S1 Domain adaptation through adding or selecting features

Some DA methods discover the best way to transfer between domains by adding, deleting, or otherwise weighting features from each domain differently. In the following subsections, we describe some representative methods in detail.

### S1.1 Adding features: feature augmentation / feature replication

The *feature augmentation* or *feature replication* strategy aggregates and transforms the source and target domain features together into an augmented feature space for use during model training. For example, Frustratingly Easy Adaptation (FEDA)(*125*) maps the augmented source and target feature space by duplicating the features into three vectors, $\chi_s = (X_s, X_s, 0)$ and $\chi_t = (X_t, 0, X_t)$, to regulate the trade-off between source/target and general weights for the classifier to learn. Daumé and colleagues(*155*) provide a semi-supervised extension of the FEDA method by forcing the source and target domains to agree on the unlabelled data to leverage unlabeled data in the target domain for model training.

Although relatively easy to implement, one should be cautious when applying FEDA methods to small-scale biological data. First, biological data have high dimensionality to begin with, so duplicating features will make the dimensionality of the data even more troublesome -- especially when multiple domains are intended to be leveraged simultaneously. Domain-independent feature reduction methods such as PCA can be applied, as was done by Schneider and colleagues(*156*), to dramatically reduce the number of features. However, this approach's performance did not substantially differ from the results of simply concatenating the source and target domains. Furthermore, FEDA methods only work with homogenous datasets, while biological data is often heterogeneous across domains. Therefore, a proposed extension of FEDA that deals with data heterogeneity by using two projection matrices $P$ and $Q$, $\chi_s = (PX_s, X_s, 0_{dt})$ and $\chi_t = (QX_t, 0_{ds}, X_t)$, has been found to be generally more useful when dealing with biological data(*8*).

### S1.2 Selecting or weighting features

*Data selection* domain adaptation strategies include methods in which the respective local geometrical structures of data in both source and target domains remain unchanged. In other words, these methods select only the informative and relevant features or samples to use in training and testing, without implementing any transformational changes to the data itself. Data selection often occurs in conjunction with and before other transformation and alignment techniques (see next section), though using it on its own can offer a beneficial and simpler approach to bridging source and target domains in some situations.

Structural correspondence learning (SCL)(*157*) is an example of this data selection strategy. SCL first defines a set of frequently occurring and diverse *pivot* features (i.e. features that behave the same way for discriminative learning in all domains) on the unlabeled data from both domains, and then estimates the pivot features' covariances with non-pivot features to learn a mapping function between source and target. Another version, a manifold-based technique(*158*), employs a method termed Statistically Invariant Sample Selection (SISS) to select landmark samples from both domains based on the pairwise Hellinger distances between the samples' distributions. In some cases, SISS has been shown to be more effective than assigning non-binary weighting to samples(*159*). Of course, the drawbacks of discarding samples that do not meet these criteria when working with small-scale biological data -- when samples are already too few -- are obvious given the discussion in the main text. However, the tradeoffs between (a) using only a few informative samples, thus risking not having enough samples, and (b) selecting all samples thus risking *negative transfer* (i.e. applying knowledge from a source domain will negatively affect the performance of the model in a target domain), should be judged on an empirical and case by case basis.

### S2 Domain adaptation through parametric transformations

Another important domain adaptation strategy concentrates on *alignment*. There are different ways to align domains, including label information or dependency structure and correspondence of features. Overall, the merit of alignment techniques is that most of them do not require that label information is available for the target domain, and some also reduce the dimensionality in a way that takes into account both the source and target. Many of these methods thus also discover a (often lower dimensional) shared (sub)space between the source and target domain, rather than a transformation that maps one domain directly onto the other. A major critique of these methods is thus that they often result in less interpretable features -- and feature interpretability is a critical objective in many biological studies. Here, we briefly discuss several parametric alignment-based approaches to DA.

(Note that domain alignment can be done in a parametric or non-parametric way, depending on how the loss function is minimized; in this section we focus specifically on parametric alignment, with nonparametric approaches to both feature selection and alignment discussed in **Section S2.2**.)

### S2.1.1 Correlation based alignment

Canonical correlation analysis (CCA)(*160*) is a classical technique used to maximize the correlation between two sets of vectors -- or in the case of DA, the correlation between the feature distributions across two domains. CCA is particularly useful in heterogeneous DA problems where the goal is to find a feature transformation to bridge the heterogeneous feature spaces via finding a subspace that is shared between two domains(*161*). However, CCA by definition finds *linear* combinations of two domains, and so cannot work where a nonlinear feature subspace is desirable(*162–164*). Fortunately, advanced implementations of CCA, such as Kernel CCA (KCCA)(*162–164*), can be employed to find non-linear combinations of shared feature representations across domains.

Another correlation-based approach, Correlation Alignment (CORAL), is a simple and fast, yet powerful, technique to minimize domain shift by aligning the second-order statistics (covariance) of two distributions(*137*). CORAL tries to minimize distances between domains using the original feature spaces rather than lower-dimensional subspaces. The first step is to remove feature correlations in the source domain (i.e. 'whitening'), and then 'recolor' the source's features by the target domain's feature covariances. The advantage of the CORAL method is that it is incredibly simple and fast -- for example, it can be implemented with four lines of code in MATLAB -- and yet is still very effective in aligning the domains. However, we note that this method is limited to aligning homogeneous domains.

### S2.1.2 Geometric transformation based alignment

Several methods assume that transformations must take on a specific functional form, potentially based on field-specific knowledge. For example, suppose one wishes to functionally align fMRI data across multiple human subjects to study shared cognitive characteristics and improve the samples-to-features ratio. Some methods to accomplish this goal use rigid-body geometric transformations because they assume that specific regions of the brain (e.g. ventral temporal cortex) encode similar features across domains, but that these features are not labeled (i.e., the coordinate system of voxels that represent specific features are not aligned). These unlabeled features across subjects must be aligned into a common feature space across multiple subjects without warping the feature space – i.e., using only translation, scaling, and rotation. Here, subjects are considered as different source domains, where the dimensionality and order of voxels (features) are different across subjects. These characteristics thus suggest a heterogeneous multi-source DA problem.

One such method, hyperalignment(*24, 165*), assumes that the shared feature space is high-dimensional. This method also assumes that data from all source domains share the same feature space that is also anchored by the same temporal variance (e.g. all domains/participants have watched the same sequence of a movie). Therefore, under the assumption that the features in each domain follow a shared trajectory in some shared feature space while anchored in time, it is possible to rotate the individual feature spaces into a common feature space. (Other variants of hyperalignment technique have been proposed(*75, 166*).) This rotation is done through Procrustean transformation – an orthogonal transformation that minimizes Euclidean distance between features across domains. This model finds hyperalignment parameters that will map each source into a common feature space of voxel responses. This method implements two main assumptions that must be considered before applying it to other datasets. First, there exists a feature space that is common between all domains, and second, there exists a linear transformation that can map the voxel pattern of multiple domains into a common feature space such that it can minimize the distance between two features in two domains. Therefore, hyperalignment might also be useful for aligning heterogeneous multi-source DA problems where these two assumptions are satisfied, especially when there is also the opportunity to exploit temporal anchoring.

### S2.1.3 Other parametric transformations

Probability distributions lie on a Riemannian manifold, and there are some alignment-based methods that exploit this fact. One method(*167*) utilizes manifold alignment without the need for any pairwise correspondence information between the source and target. More concretely, for a given source $X_s = \{x_i, \cdots, x_m\}$ of feature dimension $p$ and target $X_t = \{x_j, \cdots, x_n\}$ of feature dimension $q$, the method computes functions α and β to map source and target domains to a new lower dimensional space so that $\alpha^T x_i$ and $\beta^T x_j$ can be directly compared in order to minimize the loss function. Another similar method(*77*) expands manifold alignment to work on multiple domains with heterogeneous datasets and leverages the labels to align the domains rather than the often-inaccessible correspondence information (i.e. instances in one dataset that correspond to, or are in some way equivalent to, instances in another dataset).

Other methods can be seen as less rigid versions of the methods discussed above. For example, like hyperalignment, the Shared Response Model (SRM) was originally developed to aggregate fMRI data across many subjects to evaluate cognitive states across groups rather than within individuals(*168*). However, unlike hyperalignment, SRM projects subjects' data into a shared *lower-dimensional* feature space, and thus relaxes the assumption of rigid-body transformation and precisely shared (but unlabeled) features within a given brain region that are shared across participants. The shared space, $S$, can be calculated as follows

$$min\, W_{i,S} \sum_{i=1}^{m} \left|\left|X_i - W_i S\right|\right|^2$$

where $||.||$ is Frobenius norm, and $X_i$, $W_i$ are the fMRI responses and bases for the participant $i$, respectively. This is under the orthogonality assumption of $W_i$ such that $W_i^T W_i = I_K$ where $K$ is the dimension of shared feature space that is chosen by the experimenter.

Likewise, in the microbiome field, specific frameworks have been developed to model the processes that generate variability between different studies or batches - for example by modeling experimental variability as multiplicative bias that affects the measured taxonomic abundances(*169*). These functional forms can then form the basis for field-specific domain adaptation techniques. One such method is DEBIAS-M(*141*), a microbiome-specific domain adaptation method that realigns sources by inferring their underlying processing biases.

Likewise, in the microbiome field, specific frameworks have been developed to model the processes that generate variability between different studies or batches - for example by modeling experimental variability as multiplicative bias that affects the measured taxonomic abundances(McLaren et al. 2019). These functional forms can then form the basis for field-specific domain adaptation techniques. One such method is DEBIAS-M(Austin et al. 2024), a microbiome-specific domain adaptation method that realigns sources by inferring their underlying processing biases.

## S2.2 Domain adaptation through nonparametric feature selection and transformation approaches (neural networks)

Sometimes, it is not possible to define *a priori* the type of transformation that might be appropriate to align domains or select features. In this case, it is advantageous to turn to neural networks, which can discover and approximate any parametric function. Neural network based DA techniques thus typically have a feature extractor section in their architectures, as well as relying on standard objective functions to align domains. This section will focus on introducing various neural network architectures and the type of biological problems that they may be well–suited for.

### S2.2.1 Adversarial based

Adversarial based neural network methods usually incorporate a discriminator component in their architectures. Unlike discrepancy based methods that try to align features by minimizing a specific statistical distance measure between domains (e.g. the CORAL, hyperalignment, or SRM methods above), adversarial methods try to "fool" a discriminator until it cannot distinguish which data is coming from which distribution(*123*, *140*, *170*, *171*), and as a consequence, the network will learn domain-invariant features(*130*, *172*).

Adversarial-based methods can be separated into two categories: adversarial generative and adversarial discriminative. Adversarial generative methods, usually based on Generative Adversarial Networks (GANs), were originally designed to solve problems where there is interest in generative models(*173*). For instance, Xu and colleagues(*122*) used GAN loss in the objective function in order to minimize the discrepancy between each source domain and target by using multi-adversarial learning. Although GANs can create fascinating visualizations, they are not optimized for discriminatory tasks and are limited to domains where the shift between distributions is small(*171*). Moreover, these networks usually need many training samples, which of course is an issue in biological datasets, and hence might not be a good choice for biological data. On the other hand, adversarial discriminative methods aim to mitigate the negative effects of domain shift by learning a discriminative representation of the source and target domains(*123*, *171*) without the need for a generative component. These methods, however, have also typically been used on image data where there are tens of thousands of examples available, and so require further examination to explore whether they can be suitable for small biological datasets.

To utilize these methods, however, one has to use a feature extractor section in the architecture of the model in order to map the original input spaces into a (usually) lower-dimensional space. A simple yet effective architecture of adversarial-based DA is the domain-adversarial neural network (DANN)(*114*), developed by Ganin and colleagues, which uses a gradient reversal in order to maximize the loss of a categorical classifier and hence to ensure that features in the two domains are similar.

### S2.2.2 Autoencoder based

Autoencoder (AE) networks are unsupervised learning algorithms that consist of an encoding and a decoding section, which enable them to learn hidden representations of an input(*174*). The encoding section of an autoencoder usually uses a nonlinear function in order to discover features in a bottleneck (the dimensionality of the bottleneck is usually lower than the initial input) that are informative enough to be used by the decoding section of the network to reconstruct the original input. In DA, it is possible to learn domain invariant features by sharing the same encoding section across multiple domains. The unsupervised nature of these AE networks makes them a good candidate to discover domain-invariant features even in the case of unlabeled or sparsely labeled source domains.

Previously, AEs have been successfully used to extract domain-invariant features. For instance, stacked denoising autoencoders (SDAEs) have been used before to extract high-level features that are common across source and target domains(*175*). Therefore, a classifier trained on these high-level common features using the labels available in the source domain can also perform well on the target domain. Similarly, Bousmalis and colleagues(*176*) developed the Domain Separation Network (DSN) using two encoding sections. First, a shared encoding network between source and target domains learns the shared representation across domains, and second a private encoder learns domain-specific features. Then the decoding section will use both domain-invariant and domain-specific representations to reconstruct the input samples. Finally, the authors trained a classifier only on the shared representation so the classifier can also perform well on the target domain. As another example, Ponimova and colleagues(*90*) successfully developed the Fader network, which used a convolutional autoencoder to address the heterogeneous nature of multi-site fMRI data to increase the sample size and extract domain-invariant features across multiple subjects. Although the total number of samples were 1000, they were collected from 19 sites which means the average available data from each site was 52 samples. This success offers hope for implementing deep learning techniques with other sorts of biological datasets where the sample size from each source is limited but combining many sources can increase the overall sample size and thus prevent overfitting while discovering domain-invariant features.

### S2.2.3 Convolutional neural network based

Convolutional neural networks (CNNs)(*47*) are great candidates for DA on data that contain spatial information such as image data. In CNNs, once features are learned using convolutional and pooling operations, the exact position of these features becomes less relevant. To address the variability of feature positions between individuals, CNNs often utilize pooling operations where the receptive field (i.e. regions in the image that the CNN's feature detector can see) is subsampled or downsampled.

Previously, many have used CCN based methods such as AlexNet(*44*) and ResNet(*177*) as the backbone in the architecture of feature extractor section in DA techniques in the field of medical imaging and computer vision(*138*, *178*). For example, Chen and colleagues(*179*) developed a novel technique using a multi-view convolutional autoencoder, which combined latent variables and searchlight-based analysis, to align fMRI data from multiple human subjects and map those subjects' data into a common space. The technique was able to preserve the spatial locality of the voxels, showing comparable or superior decoding accuracy compared to other standard techniques such as SRM or standard searchlight analysis(*179*).

### S2.2.4 Recurrent neural network based

Recurrent neural networks (RNNs) contain one or more loops in their directed connection graph through which their internal state will evolve over time in discrete steps. RNNs are often used to process time-series data such as video or text, in which a new frame or character is fed into the network at each time step. Essentially, RNNs resemble Artificial Neural Networks (ANNs) with hidden layers that feed back onto themselves. This allows the layers to not only pass on the information to the next layer but also update their own weights and all the weights in previous layers. In other words, RNNs are many copies of the same ANNs that pass outputs to a successor. This makes RNNs great networks that can model phenomena which happen in a sequence of actions making them good candidates to extract features from times series data such as decoding emotions from EEG(*180*), movement control of prosthetics using electromyography (EMG)(*181*), or protein-protein interactions(*182*)(*183*). RNNs can thus be used in DA approaches: For instance, Sonsil(*181*) and colleagues developed a method where the EMG data from one individual (source domain) was used to train a neural network to predict the gestures of another individual (target domain). They used a combination of RNN and adversarial domain adaptation (ADA)(*184*) by summing up the loss functions of the RNN predictor and the discriminator.

## S3 Discovering domain-invariant spaces: a sampling of current algorithms

Having introduced classifications and useful vocabulary both above and in the main text, we also want to direct the reader to a usefully curated corner of the DA literature. In this section, we focus on DA methods that aim to discover domain-invariant (often lower dimensional) spaces of features, as these kinds of spaces may be argued to best serve the goal of discovering generalizable truths in biology.

To discover domain-invariant spaces, we aim to find a projection matrix that minimizes statistical distance between two domains (e.g. the Maximum Mean Discrepancy (MMD)). For instance, the Distribution Matching Embedding (DME)(*185*) aims to find a projection matrix, $W$, that minimizes MMD such that $W$ remains orthogonal (i.e. $W^T W = I$)). An alternative approach is to learn the common feature space using deep learning models (i.e. deep domain adaptation). However, these methods usually require a large number of samples for training. There have been some deep domain adaptation methods that have been successful on relatively smaller datasets, such as the Office-31(*53*) dataset (with ~ 132 samples per category form 3 domains), using with AlexNet(*44*) or ResNet-50(*177*) backbones(*121*, *139*, *186*, *187*). However, it is crucial to note that: (1) these are image datasets, which (2) contain substantially more samples compared to many biological datasets despite their relative sparseness compared to standard benchmark datasets. Therefore, careful consideration is necessary when applying deep DA on biological datasets.

**Table S1** describes a summary of the subcategories of domain-invariant-feature based methods along with their limitations and strengths. These approaches and their kin can ideally be used, and further developed, to help discover generalizable truths in biological data which are not limited by the idiosyncrasies of a few small datasets.

| Alignment type | Limitations | Benefits | Sample methods / citations |
| --- | --- | --- | --- |
| Shallow DA: discrepancy based (e.g. MMD(*188*), correlation alignment (CORAL)(*137*), Contrastive domain discrepancy (CCD)(*189*) | Is not assumption-free, limiting its learning ability in finding domain-invariant features. | Requires relatively less amount of data. | M3SDA(*46*) Guo et al.(*138*) Zhu et al.(*121*) Hoffman et al.(*178*) Guo et al.(*190*) |
| Deep DA: discrepancy based | Requires relatively more samples; some metrics such as MMD and CORAL might require assumptions of kernel function for calculating the distance between distributions. | Can find highly non-linear relationships between domains and the common feature space-the alignment part do not introduce new parameters. | Long et al.(*54*) Deep CORAL(*115*) Contastive Domain Adaptation (CDA)(*189*) Deep-JDOT(*191*) |
| Reconstruction based | Requires relatively more samples. | Assumption free; can find non-linear relationships between domains and the common feature space; does not need a distance metric. | DRCN(*192*) DSN(*176*) MTAE(*193*) |
| Adversarial based (e.g. GAN loss, H-divergence, Wasserstein distance(*191*)) | Requires many samples, since discriminator is a network that introduces more parameters to be learned. | Effective for medical imaging data such as MRI, CT. | ADDA(*171*) Tsai et al.(*194*) DANN(*96*) CycleGAN(*195*) CyckeEmotionGAN(*196*) |

**Table S1. Sample algorithms used in DA which seek domain-invariant spaces, shared across two or more domains, by finding a projection that minimizes the discrepancy between domains.**